\documentstyle[prl,aps,preprint,tighten]{revtex}
\def\D{{\cal D}}
\def\d{{\partial}}
\def\r{{\bf r}}
\def\x{{\bf x}}
\def\y{{\bf y}}
\def\k{{\bf k}}
\def\umin{{u_{\text{min}}}}
\def\mean#1{{\langle #1 \rangle}}
\def\ket#1{{| #1 \rangle}}

\begin{document}
\setcounter{page}{0}
\def\footnoterule{\kern-3pt \hrule width\hsize \kern3pt}

\title{Turbulent decay of a passive scalar in the Batchelor limit:\\ 
Exact results from a quantum-mechanical approach\thanks
{This work is supported in part by funds provided by the U.S.
Department of Energy (D.O.E.) under cooperative 
research agreement \#DF-FC02-94ER40818.}}
\author{D.~T.~Son\footnote{Email address: {\tt son@ctp.mit.edu}}}
\address{Center for Theoretical Physics \\
Laboratory for Nuclear Science \\
and Department of Physics \\
Massachusetts Institute of Technology \\
Cambridge, Massachusetts 02139 \\
{~}}

\date{MIT-CTP-2753,~ physics/9806047. {~~~~~} June 1998, revised January
1999}
\maketitle

\begin{abstract}

We show that the decay of a passive scalar $\theta$ advected by a random
incompressible flow with zero correlation time in the Batchelor limit can
be mapped exactly to a certain quantum-mechanical system with a finite
number of degrees of freedom.  The Schr\"odinger equation is derived and
its solution is analyzed for the case where, at the beginning, the scalar
has Gaussian statistics with correlation function of the form
$e^{-|x-y|^2}$.  Any equal-time correlation function of the scalar can be
expressed via the solution to the Schr\"odinger equation in a closed
algebraic form.  We find that the scalar is intermittent during its decay
and the average of $|\theta|^\alpha$ (assuming zero mean value of
$\theta$) falls as $e^{-\gamma_\alpha Dt}$ at large $t$, where $D$ is a
parameter of the flow, $\gamma_\alpha={1\over4}\alpha(6-\alpha)$ for
$0<\alpha<3$, and $\gamma_\alpha={9\over4}$ for $\alpha\geq3$, independent
of $\alpha$.

\end{abstract}
\vspace*{\fill}

Kolmogorov theory (K41) \cite{K41} remains the cornerstone of our
understanding of fully developed turbulence.  This simple theory predicts
a scaling law (the famous Kolmogorov-Obukhov $k^{-5/3}$ law) of the energy
spectrum that is in remarkable agreement with experimental data. Since the
1980's, however, data gathered have consistently pointed out the failure
of K41 in predicting the scaling law of high-order correlation functions
\cite{Anselmet,Frisch}.  The breakdown of K41 is closely related to the
non-Gaussianity of the distribution of velocity increments. The
phenomenon, dubbed intermittency, has become one of the central issues of
theoretical works on turbulence.  Recently, it has been found that the
intermittency of a passive scalar advected by a turbulent flow might be
even stronger than that for the velocity \cite{Sreenivasan}.  Such
observations have led to the hope that the study of simple models, such as
the Kraichnan model of scalar advection (see Refs.\
\cite{Kraichnan68,Kraichnan,Chert,GawKup,SS94,SSFr96} and below), may
provide clues to understand the much more complex Navier-Stokes
intermittency.

In this paper, we consider the problem of turbulent decay of a passive
scalar.  In other words, we want to find statistical properties of a
scalar $\theta$ satisfying the equation
\begin{equation}
  \d_t\theta + v_i\d_i\theta = \kappa\Delta\theta,
  \label{advection}
\end{equation}
where $\kappa$ is a small diffusivity, $v_i$ is a Gaussian random field,
which is white in time,
\begin{equation}
  \mean{v_i(t,\x) v_j(t',\y)}= \delta(t-t') f_{ij}(\r),
  \label{vv}
\end{equation}
and
\begin{equation}
  f_{ij}(\r) = V\delta_{ij} - D\biggl( 
  {\xi+2\over\xi}\delta_{ij}r^\xi - r^{\xi-2}r_ir_j \biggr),
  \label{fVD}
\end{equation}
where $\r=\x-\y$, and $\xi$ is some real number. The Kraichnan model
usually contains a random external scalar source in the right-hand side
(RHS) of Eq.\ (\ref{advection}).  Such a source would make the steady
state possible, but since we are interested in the decay, it is assumed
that the source is absent.  We will, furthermore, turn our attention to
the Batchelor limit $\xi=2$, which corresponds to smooth flows with very
large velocity correlation lengths (for comparison, the inertial range of
real turbulence corresponds to $\xi={2\over3}$.)  This limit has attracted
recent interest due to its good analytical features \cite{Migdal,Bernard}.

Our result is that the scalar becomes more and more intermittent during
the decay.  Specifically, we found that the average of
$\mean{|\theta(x)|^\alpha}$, where $\alpha$ is an arbitrary positive
number, decays as $e^{-\gamma_\alpha Dt}$ at asymptotically large $t$,
where $\gamma_\alpha={1\over4}\alpha(6-\alpha)$ if $\alpha<3$ and
$\gamma_\alpha={9\over4}$ when $\alpha\geq3$.  The flatness
$\mean{\theta^4}/\mean{\theta^2}\sim e^{7Dt/4}$ goes to $\infty$ as $t$
grows.  This is in sharp contrast with the steady-state case, where the
scalar statistics is largely Gaussian \cite{Bernard}.

To attack the problem, we will reduce it to a certain problem of quantum
mechanics, which can then be solved (for another attempt to apply quantum
mechanics to turbulence, see \cite{MM96}.) We first note that the
probability distribution functional of the scalar, which will be denoted
$\Psi[t,\theta]$, can be expressed in term of a path integral
\cite{MSR}
\begin{equation}
  \Psi[t,\theta] = \int\!\D\pi(t,\x)\,\D\theta(t,\x)\,\D v_i(t,\x)\, 
  \rho[v]\,\exp\biggl[i\int\!dt\,d{\x}\,\pi(\d_t\theta+v_i\d_i\theta-
  \kappa\Delta\theta)\biggr],
  \label{pathint}
\end{equation}
where the Gaussian measure for the velocity $\rho[v]$ is chosen to satisfy
Eq.\ (\ref{vv}).  The auxiliary variable $\pi$ enforces Eq.\ 
(\ref{advection}).  Integrating over $v$, one obtains
\begin{eqnarray*}
  \Psi(t,\theta) & = &
  \int\!\D\pi\,\D\theta\, \exp\biggl[i\int\!dx\,\pi\d_t\theta
  -{1\over2} \int\! dt\, d\x\, d\y\, 
  \pi(t,\x)\d_i\theta(t,\x)f_{ij}(\x-\y)\pi(t,\y)\d_j\theta(t,\y) \\
  & & - i\kappa\!\int\! dx\, \pi\Delta\theta\biggr].
\end{eqnarray*}
The path integral describes the evolution in Euclidean time of a quantum
field theory with the Hamiltonian \cite{Slavnov}
\begin{equation}
  H = {1\over2} \int\! d\x\, d\y\, \pi(\x)\d_i\theta(\x)
      f_{ij}(\x-\y)\pi(\y)\d_j\theta(\y)
      + i\kappa\!\int\! dx\, \pi\Delta\theta,
  \label{Hamiltonian}
\end{equation}
where $\theta$ and $\pi$ are conjugate variables satisfying the usual
commutation relation $[\theta(\x), \pi(\y)]=i\delta(\x-\y)$.  The operator
ordering in Eq.\ (\ref{Hamiltonian}) corresponds to the physical
regularization of the path integral (\ref{pathint}).  The evolution of the
distribution functional $\Psi[\theta]$ is described by the Euclidean
version of the Schr\"odinger equation, $\d_t\Psi = - H\Psi$. Note that the
functional $\Psi$ itself, not its square, determines the probability
distribution of $\theta$.  The average of, e.g., $|\theta|^\alpha$ is
defined as $\mean{|\theta|^\alpha}=\int\!{\cal
D}\theta\,|\theta|^\alpha\Psi[\theta]$. In further discussion, we will use
the quantum-mechanical terminology, so the terms ``probability
distribution functional'' (PDF) and ``wave function'' are used
interchangeably.

In the Batchelor limit (\ref{fVD}), the Hamiltonian can be simplified
considerably. We will concentrate our attention to the homogeneous case,
i.e., when the system is invariant under spatial translations.  In the
quantum language, this means that we restrict ourselves to the states
$\ket{\Psi}$ having zero total momentum, $P_i\ket{\Psi}=0$, where $P_i =
\int\! d\x\,\pi(\x)\d_i\theta(\x)$ \cite{Itzykson}.  With this
restriction, the Hamiltonian (\ref{Hamiltonian}) can be rewritten into the
following form:
\begin{equation}
  H = {D\over2}(4L_{ij}L_{ij} - L_{ii}L_{jj} - L_{ij}L_{ji})
      + i\kappa D_{ii}
  \label{HLL}
\end{equation}
where the operators $L_{ij}$ and $D_{ij}$ are defined as 
\[
  L_{ij} = \int\! d\x\, x_i\pi(\x)\d_j\theta(\x), \quad
  D_{ij} = \int\! d\x\, \pi(\x)\d_i\d_j\theta(\x).
\]

It is straightforward to check that $L_{ij}$ and $D_{ij}$ form a closed
algebra with the commutation relations,
\begin{eqnarray}
  \left[ L_{ij}, L_{kl} \right] & = & 
    i(\delta_{jk}L_{il} - \delta_{li}L_{kj}), \nonumber \\
  \left[L_{ij}, D_{kl}\right] & = & -i(\delta_{il}D_{jk} +
    \delta_{ik}D_{jl}), \label{LDcomm} \\
  \left[D_{ij}, D_{kl}\right] & = & 0. \nonumber 
\end{eqnarray}
The fact that the algebra is closed implies that the system is actually
one with a finite number of degrees of freedom.  The quantum field theory
thus degenerates to quantum mechanics.  Notice that $L_{ij}$ form a closed
subalgebra.  Indeed, they are the operators of linear coordinate
transformations. In fact, only the SL(3,R) generators enter the
Hamiltonian (\ref{HLL}) (cf.\ \cite{SSFr96}.)  $H$ is invariant under the
SO(3) algebra formed by the antisymmetric part of $L_{ij}$.

In principle, the Schr\"odinger equation with $H$ defined in Eq.\
(\ref{HLL}) can be solved (at least numerically.)  In this paper, we will
choose a representation of the algebra (\ref{LDcomm}) where $H$ has a
relatively simple form, but the physics is nontrivial.  Our choice is
inspired by the observation by Townsend \cite{Townsend} that a Gaussian
shape hot spot preserves its Gaussianity when advected by Batchelor-limit
velocity flow (for a somewhat similar discussion without quantum
mechanics, see \cite{KK93}.) Let us for a moment concentrate on the states
in which $\theta$ has Gaussian statistics. This corresponds to the wave
functions of the form $\Psi[\theta]\sim\exp(-{1\over2}\theta
K^{-1}\theta)$, where $K(x-y)=\mean{\theta(x)\theta(y)}$.  We will further
restrict ourselves on functions $K$ that have the Gaussian shape,
$K(x-y)\sim\exp[-{1\over2}b_{ij}(x-y)_i(x-y)_j]$.  More strictly, we
require that, in Fourier components, the spectrum of $\theta$ has the form
\[
  \mean{\theta^*(\k)\theta(\k')}=\theta_0
  \exp\biggl(-{1\over2}a_{ij}k_ik_j\biggr) \delta(\k-\k'),
\]
where $\theta_0$ is a constant independent of $a_{ij}=(b_{ij})^{-1}$ (one
can choose $\theta_0=1$.)  Denote such states as $\ket{a_{ij}}$.  The
group elements act on $\ket{a_{ij}}$ as follows:
\begin{eqnarray}
  e^{-i\beta_{ij}L_{ij}} \ket{a_{ij}} & = &
  \ket{e^{-\beta}a(e^{-\beta})^T} \quad 
  \text{if} \quad \beta_{ii}=0 \nonumber\\
  e^{-i\beta_{ij}D_{ij}} \ket{a_{ij}} & = & \ket{a_{ij}+4\beta_{ij}}.
  \label{group_rep}
\end{eqnarray}

We now choose our representation to be the one acting on the subspace of
the Hilbert space that contains all linear combinations of $\ket{a_{ij}}$
(although the latter do not form an orthogonal basis.)  A vector in this
subspace is characterized by the function $\psi(a_{ij})$, which is the
coefficient of the expansion
$\ket{\Psi}=\int\!da_{ij}\,\psi(a_{ij})\ket{a_{ij}}$.  In general, the
scalar statistics in $\ket{\Psi}$ is not Gaussian.  The operators $L_{ij}$
and $D_{ij}$ can be written as first-order differential operators with
respect to $a_{ij}$, and the Schr\"odinger equation becomes a second-order
PDE on $\psi$. 

Moreover, if the initial condition is isotropic, i.e., invariant under
SO(3) rotations $\epsilon_{ijk}L_{jk}$, the wave function depends only on
the eigenvalues of the matrix $a_{ij}$, not on the Eulerian angles
characterizing the orientation of the eigenvectors.  The wave function is
now a function of three variables, $\psi(u_1,u_2,u_3)$, where we have
denoted the eigenvalues of $a_{ij}$ as $e^{2u_i}$.  We re-scale $\psi$ so
that the state $\ket{\Psi}$ is expressed via $\psi(u)$ as
\begin{equation}
  |\Psi\rangle=\int\!du_i\,dU\,\psi(u)|a(u,U)\rangle,
  \label{wavefunction}
\end{equation}
where $a(u,U) = U\text{diag}(e^{2u_i})U^{-1}$, $U$ belongs to SO(3), and
the integration over $U$ is performed using the invariant measure on the
SO(3) group manifold. 

The Schr\"odinger equation $\psi(u)$ can then be derived (details are
found in \cite{Son}). It has the form
\begin{equation}
  \d_t\psi = D(\d_1^2+\d_2^2+\d_3^2-\d_1\d_2-\d_2\d_3-\d_3\d_1)\psi-
  \sum_{i=1}^3\,
  \bigl[3D\d_i(f_i\psi) + 2\kappa\d_i(e^{-2u_i}\psi)\bigr],
  \label{Schroedinger}
\end{equation}
where $\d_i\equiv\d/\d u_i$, 
\begin{eqnarray}
  f_1 & \equiv & f(u_1;u_2,u_3) =
  {e^{4u_1}-e^{2(u_2+u_3)} \over
  (e^{2u_1}-e^{2u_2})(e^{2u_1}-e^{2u_3})}\, ,
  \nonumber \\
  f_2 & \equiv & f(u_2;u_3,u_1), \quad  f_3\equiv f(u_3;u_1,u_2).
  \label{f} 
\end{eqnarray} 
Special caution is required when two of $u_i$ are equal to each other,
however this will not affect our subsequent discussion. 

To fully define the problem, the initial condition of $\psi(u)$ is needed. 
One can take as the initial state the vector $\ket{a_{ij}}$, where
$a_{ij}=\text{diag}(1,1,1)$.  This corresponds to a scalar that has
Gaussian statistics, zero mean value, and the correlation function
$\mean{\theta(x)\theta(0)}$ proportional to $e^{-x^2/2}$ at $t=0$.  The
correlation length of $\theta$ is taken to be of order 1.  In terms of
$\psi$, the initial condition is
$\psi(t=0,u)=\delta(u_1)\delta(u_2)\delta(u_3)$. 

Equation (\ref{Schroedinger}) can be interpreted in an intuitive way by
using a three-dimensional random walk that has the Fokker-Planck equation
coinciding with Eq.\ (\ref{Schroedinger}) \cite{Risken},
\begin{equation}
  \dot{u}_i = 3D f_i + 2\kappa e^{-2u_i} + \xi_i,
  \label{Langevin}
\end{equation}
where $\xi_i$ are white noises that correlate as follows:
\begin{eqnarray}
  & & \xi_1+\xi_2+\xi_3 = 0,
  \nonumber \\
  & & \mean{\xi_1(t)\xi_1(t')} = \mean{\xi_2(t)\xi_2(t')} 
      = \mean{\xi_3(t)\xi_3(t')} = 2D\delta(t-t'),
  \label{sumxi}\\
  & & \mean{\xi_1(t)\xi_2(t')} = \mean{\xi_2(t)\xi_3(t')} =
       \mean{\xi_3(t)\xi_1(t')} = -D\delta(t-t').
  \nonumber
\end{eqnarray}
Let us discuss the physical meaning of Eq.\ (\ref{Langevin}).  A point
$(u_1,u_2,u_3)$ corresponds to the configuration of $\theta$ having the
spectrum $\mean{|\theta(\k)|^2}\sim\exp(-{1\over2}\sum e^{2u_i}k_i^2)$. In
the configuration space, $\theta$ is approximately constant inside an
ellipsoid with major axes proportional to $e^{u_i}$.  When advected by the
flow, this ellipsoid is subjected to random linear transformations.  If
the only transformations of the ellipsoids are those that stretch or
compress the ellipsoid in the directions of its major axes, the results
would be $\dot{u}_i=\xi_i$, where $\xi_i$ are random.  Equation
(\ref{sumxi}) reflects the conservation of the volume of the ellipsoid
during random stretching and compressing.  However, the ellipsoid may be
subjected to stretching or compressing in directions other than the major
axes, as well as to shearing.  These effects are accounted for by the term
$3Df_i$ on the RHS of Eq.\ (\ref{Langevin}).  The incompressibility is not
violated, due to the identity $f_1+f_2+f_3=0$.  The terms $2\kappa
e^{-2u_i}$ are not important unless one major axis of the ellipsoid is as
small as the diffusion scale.  In the latter case, diffusion smears out
the scalar and causes it to be correlated at a larger distance.  This is
exactly the effect of the $2\kappa e^{-2u_i}$ terms in the Langevin
equation.  Due to the sign of these terms, the volume of the ellipsoid
and, hence, also $u_1+u_2+u_3$, always grows during the random walk.

Since any correlation function can be computed for $|a_{ij}\rangle$, where
the scalar statistics is Gaussian, one can find any correlation function
with respect to $\ket{\Psi}$ if one knows the solution to Eq.\
(\ref{Schroedinger}) (e.g., from numerical integration.) For example, the
average of $|\theta|^\alpha$ ($\alpha>0$) over the state $\ket{a(u,U)}$ is
proportional to $e^{-\alpha(u_1+u_2+u_3)/2}$; therefore, its average with
respect to $\ket{\Psi}$ is,
\[
  \mean{|\theta|^\alpha} = C_\alpha \mean{\theta^2(t=0)}^{\alpha/2}\!
  \int\!du\,\psi(u)\exp\biggl[-{\alpha\over2}(u_1+u_2+u_3)\biggr],
\]
where $C_\alpha=\pi^{-1/2}2^{\alpha/2}\Gamma[(\alpha+1)/2]$.  This
relation is exact.

When $\kappa$ is small, the exponential behavior of
$\mean{|\theta|^\alpha}$ can be found analytically.  This can be done by
using the path-integral description of the random walk (\ref{Langevin})
and finding the saddle-point trajectories that dominate $|\theta|^\alpha$
\cite{Son}.  In this paper, we use a heuristic, yet more physical, method
to find the large time behavior of $\mean{|\theta|^\alpha}$. 

Let us assume that after letting the system (\ref{Langevin}) evolve for a
while, the values of $u_1$, $u_2$, and $u_3$ become widely separated.  We
assume $u_1<u_2<u_3$, and wide separation means $u_2-u_1\gg1$,
$u_3-u_2\gg1$.  From Eq.\ (\ref{f}) one sees immediately that in this
regime $f_1=-1$, $f_2=0$, and $f_3=1$ (in fact, these asymptotic values of
$f_i$ are related to the Lyapunov exponents, see, e.g.,
Ref.\cite{GambaKolokolov}).

Let us first ignore the term proportional to diffusivity in Eq.\
(\ref{Langevin}).  The velocity $\dot{u}_i$ has two contributions: one
from $f_i$ and another from the noise $\xi_i$.  The first contribution
implies that the mean values of $u_i$ drift with constant velocities,
$u_1(t)=-3Dt$, $u_2(t)=0$, and $u_3(t)=3Dt$, while the noises make $u_i$
fluctuate around these mean values.  The condition of wide separation of
$u$'s is satisfied when $t\gg D^{-1}$.  The advection, on average,
compresses a fluid element in one direction by a factor of $e^{3Dt}$ and
stretches it in another direction by the same factor.  The remaining third
direction is not substantially compressed or stretched. In this regime,
the diffusion is still not operative, and $\mean{|\theta|^\alpha}$ remains
constant.

At $t=(6D)^{-1}\ln\kappa^{-1}$ ($\gg D^{-1}$ if $\kappa$ is very small),
the mean value of $u_1$ becomes ${1\over2}\ln\kappa$. The term $\kappa
e^{-2u_1}$ in the Langevin equation (\ref{Langevin})  cannot be ignored
anymore.  Physically, regions of different $\theta$ have been brought this
close together so that diffusion is no longer negligible.  Let us consider
the equation for $u_1$, $\dot{u}_1=-3D+2\kappa e^{-2u_1}+\xi_1$, near
$\umin={1\over2}\ln\kappa$. The first term on the RHS pushes $u_1$ toward
smaller values, while the second term prevents $u_1$ from becoming
substantially smaller than $\umin$.  The variable $u_1$ thus fluctuates
around $\umin$.  Therefore, the random walk becomes effectively
two-dimensional:
\begin{eqnarray}
  & & \dot{u}_2 = \xi_2,\quad \dot{u}_3 = 3D + \xi_3, 
  \nonumber \\
  & & \mean{\xi_2(t)\xi_2(t')} = \mean{\xi_3(t)\xi_3(t')} 
      = 2D\delta(t-t'), \label{randwalk2} \\
  & &  \mean{\xi_2(t)\xi_3(t')} =  -D\delta(t-t'). \nonumber
\end{eqnarray}

Additionally, it is required that $u_2+u_3$ not decrease with time, due to
the previously found fact that $u_1+u_2+u_3$ can only increase (if
$u_2+u_3$ decreases, this means that $u_1$ steps away from the value
$u_1=\umin$.) Now there is a possibility for $|\theta|^\alpha$ to decay,
since it is proportional to $e^{-\alpha(u_1+u_2+u_3)/2}$, but
$u_1+u_2+u_3$ is no longer a constant.  Assuming that the random walk
(\ref{randwalk2})  starts at $u_2=u_2^0$ and $u_3=u_3^0$, the distribution
of $u_2$ and $u_3$ at large times is Gaussian: 
\begin{equation}
  \rho(u_2,u_3) \sim \exp\biggl\{ -{1\over3Dt}
  \biggl[(u_2-u_2^0)^2+(u_3-u_3^0-3Dt)^2 +
  (u_2-u_2^0) (u_3-u_3^0-3Dt)\biggr] \biggr\}
  \label{distr2}
\end{equation}
The mean value of $|\theta|^\alpha$ can be computed by taking the average
of $e^{-\alpha(u_2+u_3)/2}$ over the distribution (\ref{distr2}). 
Consider the case of $0<\alpha\leq3$ first.  The integral
$\int\!du_2\,du_3\, \rho(u_2,u_3)e^{-\alpha(u_2+u_3)/2}$ is dominated by
the region near $u_2-u_2^0=-{1\over2}\alpha Dt$,
$u_3-u_3^0=(3-\alpha/2)Dt$.  The value of the average is proportional to
$e^{-\gamma_\alpha Dt}$, where $\gamma_\alpha={1\over4}\alpha(6-\alpha)$. 

Note that the region where the integral is saturated has $u_2$ decreasing
with time, $u_2=u_2^0-{1\over2}\alpha Dt$.  Eventually, $u_2$ will become
as small as $\umin$, and the term $\kappa e^{-2u_2}$ in the Langevin
equation becomes important.  Now, both $u_1$ and $u_2$ fluctuate around
$\umin$.  However, as we will explain, the exponential decay law does not
change. Indeed, when $u_1$ and $u_2$ remain approximately constant, the
evolution of $u_3$ is described by the one-dimensional random walk,
\[
  \dot{u}_3 = 3D + \xi_3, \quad
  \mean{\xi_3(t)\xi_3(t')} = 2D\delta(t-t').
\]
The distribution of $u_3$ is now
$\rho(u_3)\sim\exp[-(4Dt)^{-1}(u_3-u_2^0-3Dt)^2]$.  Taking the average of
$e^{-\alpha u_3/2}$ (which is proportional to $\mean{|\theta|^\alpha}$
since $u_1$ and $u_2$ are constant), one finds that the decay law is still
$e^{-\gamma_\alpha Dt}$ where $\gamma_\alpha={1\over4}\alpha(6-\alpha)$.

For the particular case $\alpha=2$, our result can be checked against the
calculations based on the exact evolution equation for the scalar spectrum
\cite{Kraichnan68}.  This comparison has been done; the results indeed
agree.

When $\alpha>3$, the solution $u_2\sim-{1\over2}\alpha Dt$,
$u_3\sim(3-{\alpha/2})Dt$ is no longer realizable, since it has
decreasing $u_2+u_3$.  The average of $|\theta|^\alpha$ is then determined
by the edge of the distribution function, i.e., by $u_2\sim-{3\over2}Dt$
and $u_3\sim{3\over2}Dt$, or, after $u_2$ reaches $\umin$,
$u_2\approx\umin$ and $u_3\sim\text{const}$.  The expectation value decays
as $e^{-9Dt/4}$.  The reason the decay law does not contain $\alpha$
is the following: when $\alpha\geq3$, the main contribution to
$|\theta|^\alpha$ comes from the realizations in the statistical ensemble
where $\theta$ is unaffected by diffusion (i.e., the ellipsoid in which
$\theta$ is approximately constant has never been too thin during its
evolution.) The average $\mean{|\theta|^\alpha}$ is thus determined by
the probability of such realizations, which depends only on
characteristics of the flow but not on $\alpha$.  This probability, as has
been found, falls as $e^{-9Dt/4}$.  This implies, in particular, that the
flatness $\mean{\theta^4}/\mean{\theta^2}$ grows as $e^{7Dt/4}$, meaning
that the scalar becomes more and more intermittent during its decay.

More careful analysis shows that the decay law $e^{-\gamma_\alpha Dt}$
that we have found is valid only at large enough $t$.  At intermediate
$t$, there is a smooth transition from
$\mean{|\theta|^\alpha}=\text{const}$ to $\mean{|\theta|^\alpha}\sim
e^{-\gamma_\alpha Dt}$ \cite{Son}.  The full analysis does not change the
long-time tail of $\mean{|\theta|^\alpha}$. 

In conclusion, we have shown that by mapping to quantum mechanics, the
problem of turbulent decay of a randomly advected scalar in the Batchelor
limit can be made completely solvable.  The power of the approach
described in this paper is not limited to the calculations of
$\mean{|\theta|^\alpha}$; analogous calculations can be done for any
equal-time correlation function.  For example, the long-time tail of
$\mean{|\partial_x\theta|^\alpha}$ is also $e^{-\gamma_\alpha Dt}$ with
the same $\gamma_\alpha$.  The situation here is not similar to the steady
state, where the scalar and its derivatives have very different
statistics, with the scalar being largely Gaussian and its derivatives
being intermittent \cite{Bernard}.  The relevance of the techniques
presented and results to the general problem of intermittency is yet to be
explored.\bigskip 

The author thanks E.~Farhi, J.~Goldstone, and K.~Rajagopal for helpful
discussions, and R.~Kraichnan for pointing out Ref.\ \cite{KK93} to him.
This work was supported in part by funds provided by the U.S. Department
of Energy (DOE) under cooperative research agreement No.\
DF-FC02-94ER40818.

\end{document}